\documentclass[a4paper,twocolumn,
english,aps,pre,floatfix,showpacs]{revtex4}
\usepackage[T1]{fontenc}
\usepackage[latin1]{inputenc}
\usepackage{amsmath}
\usepackage{babel}
\usepackage{graphics}
\usepackage{amssymb}

\begin{document}
 
\title{Search for universal roughness distributions in a critical
interface model
}

\author{S.L.A. \surname{de Queiroz}}

\email{sldq@if.ufrj.br}

\affiliation{Instituto de F\'\i sica, Universidade Federal do
Rio de Janeiro, Caixa Postal 68528, 21941-972
Rio de Janeiro RJ, Brazil}

\date{\today}

\begin{abstract}
We study the probability distributions of interface roughness, sampled
among successive equilibrium configurations of a single-interface model
used for the description of Barkhausen noise in disordered magnets, 
in space dimensionalities $d=2$ and $3$. The influence of a self-regulating 
(demagnetization) mechanism is investigated, and evidence is given
to show that it is irrelevant, which implies that the model belongs to the
Edwards-Wilkinson universality class. We attempt to fit our data to the 
class of roughness distributions associated to $1/f^\alpha$ noise.
Periodic, free, ``window'', and mixed boundary conditions are examined, 
with rather distinct results as regards quality of fits to  $1/f^\alpha$  
distributions.
\end{abstract}
\maketitle
\section{INTRODUCTION}
\label{intro}

This paper deals with fluctuation properties of driven interfaces in
random media. The subject has been the focus of much current interest (for
reviews see, e.g., Refs.~\onlinecite{bar95,kar98}). Special attention has 
been given to features at and close to the depinning transition, where 
a threshold is reached for the external driving force, above which the 
interface starts moving at a finite speed. In analogy with the 
well-established scaling theory of equilibrium critical phenomena, one 
usually searches for the underlying universality classes and their  
respective critical indices, wherever such concepts are 
applicable. One example is the roughness exponent $\zeta$ which 
characterizes the disorder-averaged mean-square deviations of the 
interface about its mean height, at depinning~\cite{bar95}.  

It has been shown very recently that the probability 
distribution functions (PDFs) of critical fluctuations in seemingly 
disparate (both equilibrium and out-of-equilibrium) systems display a 
remarkable degree of universality~\cite{bhp98,bhp00,adgr01,adgr02}. 
In the context of depinning phenomena, this indicates that 
one may gain  additional insight into the physical mechanisms involved, 
by investigating the full roughness PDFs instead of concentrating on their 
lowest-order moments. 
Here we investigate the PDFs of interface roughness for a specific
single-interface model which has been used in the description of
Barkhausen noise~\cite{umm95,us,us2,bark3}, and is related to the quenched
Edwards-Wilkinson universality class~\cite{les93,ma95,mbls98,rhk03}.
A preliminary investigation of this problem was reported in
Ref.~\onlinecite{bark3}.

Barkhausen ``noise'' (BN) is an intermittent phenomenon which reflects the
dynamics of domain-wall motion in the central part of the hysteresis cycle
in ferromagnetic materials (see Ref.~\onlinecite{dz04} for an up-to-date 
review). A sample placed in a time-varying external magnetic field
undergoes sudden microscopic realignments of groups of magnetic moments, 
parallel to the field. For suitably slow driving rates, such domain-wall 
motions, or ``avalanches'', are well separated and can be easily 
individualized. The accompanying changes of magnetic flux 
are usually detected by  wrapping a coil around 
the sample and measuring the voltage pulses thus induced across the coil. 
The integral of the voltage
amplitude of a given pulse over time is proportional to the change in
sample magnetization, thus giving a measure of the number of spins
overturned in that particular event, or ``avalanche size''.
Modern experimental techniques allow direct observation, in ultra-thin 
films, of the domain-wall motion characteristic of BN, via the 
magneto-optical Kerr effect~\cite{pup00,kcs03}.

It has been proposed that BN is an illustration of ``self-organized
criticality''~\cite{bw90,cm91,obw94,umm95}, in the sense that a broad 
distribution of scales (i.e. avalanche sizes) is found within a wide range 
of variation of the external parameter, namely the applied magnetic field,
without any fine-tuning. Accordingly, the interface model studied here
incorporates a self-regulating mechanism in the form of a demagnetizing 
term (see below). In the context of interface depinning models, the 
question arises  of whether this is a relevant perturbation, i.e., whether
self-organized depinning phenomena belong to the same universality class 
as their counterparts which do not incorporate such mechanisms.

In what follows, we first recall pertinent aspects of the interface model
used here, and of our calculational methods, as well as the connections
between roughness distributions and $1/f^\alpha$ noise. Next, we 
exhibit numerical data for roughness distributions, generated by our 
simulations. We examine the influence of the self-regulating  mechanism, 
and  investigate the effect of assorted boundary conditions, both on our 
results and on the class of $1/f^\alpha$ noise distributions to which
they are compared. Finally, we discuss our
findings with regard to the relevant universality classes.

\section{Model and calculational method}
\label{sec:2}
 
The single-interface model used here was introduced in
Ref.~\onlinecite{umm95} for the description of BN. We consider the
adiabatic limit of a very slow driving rate, thus avalanches are
considered to be instantaneous (occurring at a fixed value of the external
field).

Simulations are performed on an $L_x \times L_y \times \infty$  geometry,
with the interface motion set along the infinite direction.
The interface at time $t$ is described by its height $h_i \equiv
h(x,y,t)$, where $(x,y)$ is the projection of site $i$ over the 
cross-section. No overhangs are allowed, so $h(x,y,t)$ is 
single-valued. 
We consider mainly $L_y=1$ (system dimensionality $d=2$, interface
dimensionality $d^\prime = d-1 = 1$), and $L_x=L_y$ ($d=3$,
$d^\prime=2$).
For reasons to be explained below, we will use the following sets of
boundary conditions: periodic (PBC),
so every site has two neighbors for $d=2$ and four for $d=3$; free (FBC),
meaning that the interface is horizontal at the edges ($\partial h
/\partial \widehat{n} =0$, where $\widehat{n}=\widehat{x}$ or
$\widehat{y}$ is the normal
in the cross-section plane), and mixed (MBC), i. e., periodic
along $x$ and free along $y$. These latter were employed in
Ref.~\onlinecite{bark3}, to
reproduce the physical picture of films with varying thickness. 
We also considered an alternative implementation of FBC, namely 
window boundary conditions (WBC), to be described in 
Section~\ref{subsec:fbc}.

Each element $i$ of the interface experiences a force 
given by:
\begin{equation}
f_i=u(x,y,h_i)+{k}\sum_{j} \left[h_{\ell_j(i)}- h_i\right]+H_e~,
\label{force}
\end{equation}
where
\begin{equation}
H_e =H -\eta M~.
\label{He}
\end{equation}
The first term on the RHS of Eq.~(\ref{force}) is chosen randomly, for 
each lattice site  $\vec{r_i} \equiv (x,y,h_i)$,  from a Gaussian 
distribution of zero mean and standard deviation $R$, and represents
quenched disorder. Large negative values of $u$ lead to local interface 
pinning. The second term (where the force constant $k$ is taken as the 
unit for $f$) corresponds to elastic nearest-neighbor
coupling (surface tension); $\ell_j(i)$  is the position of the $j$-th 
nearest neighbor of site $i$. For MBC, sites at $y=1$ and $y=L_y$ 
have only three neighbors on the $xy$ plane (except in the monolayer case 
$L_y=1$ which is the two-dimensional limit, where all interface sites have 
two neighbors).
The last term is the effective driving force, resulting from the applied
uniform external field $H$ and a demagnetizing field which is taken to be
proportional to
$M=(1/L_xL_y)\sum^{L_xL_y}_{i=1} h_i$,
the magnetization  (per site) of the previously flipped spins for a
lattice of transverse area $L_xL_y$.
For actual magnetic samples, the demagnetizing field is not necessarily
uniform along the sample; even when it is (e.g. for a uniformly magnetized
ellipsoid), $\eta$ would depend on the system's aspect
ratio~\cite{zcds98}. Therefore, our approach amounts to a simplification,
which is nevertheless expected
to capture the essential aspects of the problem~\cite{us2}. 
Here we use $R=5.0$, $k=1$, $\eta=0.05$, values for which fairly broad
distributions of avalanche sizes and roughness are 
obtained~\cite{us,us2,bark3}. We also  consider the effects of taking 
$\eta \equiv 0$, i.e., the non-self-organizing limit. 

We start the simulation with a flat wall. All spins above it are
unflipped. The force $f_i$ is
calculated for each unflipped site along the interface, and each spin at a
site with $f_i\geq 0$ flips, causing the interface to  move up one step.
The magnetization is updated, and this process continues, with as many
sweeps of the whole lattice as necessary, until
$f_i<0$ for all sites, when the interface comes to a halt.
The external field is then increased by the minimum amount needed to bring
the most weakly pinned  element to motion. The  avalanche size corresponds
to the number of spins flipped between two consecutive interface stops. 

On account of the demagnetization term, the effective field
$H_e$ at first rises linearly with the applied field $H$, and
then, upon further increase in $H$, saturates (apart from small
fluctuations)  at a value rather close to the
critical external field for the corresponding model {\em without}
demagnetization~\cite{umm95,us}.
The saturation $H_e$ depends on $R$, $k$ and $\eta$ 
({\em not} noticeably  on $L_x$, $L_y$)~\cite{us,bark3}, 
and can be found from small-lattice simulations. 
It takes $10^2 - 10^3$ avalanches for a steady-state regime
to be reached, as measured by the stabilization of $H_e$ against $H$.

\section{Roughness distributions and $1/f^\alpha$ noise}
\label{sec:3}
We have generated histograms of occurrence of interface roughness,
to be examined in the context of universal fluctuation 
distributions~\cite{bhp98,bhp00,adgr01,adgr02}. 
We have used only steady-state data, i.e., after the stabilization of
$H_e$ of Eq.~(\ref{He}) against external field $H$. This is the regime
in which the system  is self-regulated at the edge of 
criticality~\cite{umm95,us}. As the model
is supposed to mimic the data acquisition regime for BN, during which
the external field grows linearly in time~\cite{umm95,us,us2,bark3,dz04},
the value of $H$ is a measure of ``time''. 

At the end of each avalanche, we measured the roughness $w_2$ 
of the instantaneous interface configuration at time $t$, as 
the (position-averaged) square width of the interface
height~\cite{adgr02,rkdvw03}: 
\begin{equation}
w_2(t) 
=\left(L_xL_y\right)^{-1}\,\sum_{i=1}^{L_xL_y}\left(h_i(t)-\overline{h}(t)
\right)^2\ ,
\label{eq:rough1}
\end{equation}
where $\overline{h}(t)$ is the average interface height at $t$. As the
avalanches progress, one gets a sampling of successive equilibrium
configurations; the ensemble of such configurations yields a distribution 
of the relative frequency of occurrence of $w_2$. 
Here we usually considered ensembles of $5 \times 10^7$ events (one
and a half orders of magnitude larger than in Ref.~\onlinecite{bark3}), so
we ended up with rather clean distributions. This was essential, in
order to resolve ambiguities left over from our previous
results~\cite{bark3}.

The width distributions for correlated systems at criticality may be put
into a scaling form~\cite{forwz94,adgr01,adgr02,rkdvw03}, 
\begin{equation}
\Phi (z) = \langle w_2\rangle\,P(w_2)\ ,\quad z \equiv w_2 /\langle 
w_2 \rangle\ ,
\label{eq:pvsphi}
\end{equation}
where angular brackets stand for averages over the ensemble
of successive interface configurations, and the size dependence
appears only through the average width $\langle w_2\rangle$. By
running simulations with
${\cal O} (10^6)$ events, and $400 \leq L_x \leq 1200$ for $d=2$
$(L_y=1$),
$30 \leq L_x=L_y \leq 80$ for $d=3$~\cite{bark3}, we ascertained that
Eq.~(\ref{eq:pvsphi}) indeed holds, i.e., finite-size effects are
not detectable in any significant way as far as the scaling functions 
$\Phi (z)$ are  concerned. 
The finite-size scaling of the first
moment gives the roughness exponent~\cite{bar95}:
\begin{equation}
\langle w_2 (L)\rangle \sim L^{2\zeta}\ \ ,
\label{eq:zeta}
\end{equation}

In the context of critical fluctuation phenomena, it is known that
boundary conditions have a non-trivial effect on scaling functions, as
infinite-range critical correlations are sensitive to the boundaries of
the system~\cite{adgr01,adgr02,adr04,rkdvw03,bin81}. This is the
motivation for use of the assorted boundary conditions defined in 
Sec.~\ref{sec:2}.

We have compared our results against the family of roughness distributions 
for $1/f^\alpha$ noise, described in Refs.~\onlinecite{adgr02,rkdvw03}. 
As explained there, such distributions are derived under the assumption
that the Fourier modes into which the interface is decomposed are 
uncorrelated (generalized Gaussian 
approximation~\cite{rkdvw03}), and with amplitudes such that the 
frequency dependence of the power spectrum is purely 
$1/f^\alpha$~\cite{adgr02}. This is the simplest starting point from
which one may expect non-trivial results (the trivial ones corresponding 
to the case in which the {\em real-space fluctuations} are themselves 
uncorrelated, implying $\alpha=1/2$).  

\section{results}
\label{sec:results}

\subsection{Influence of self-regulating term}
\label{subsec:eta0}
\begin{figure}
{\centering \resizebox*{3.4in}{!}{\includegraphics*{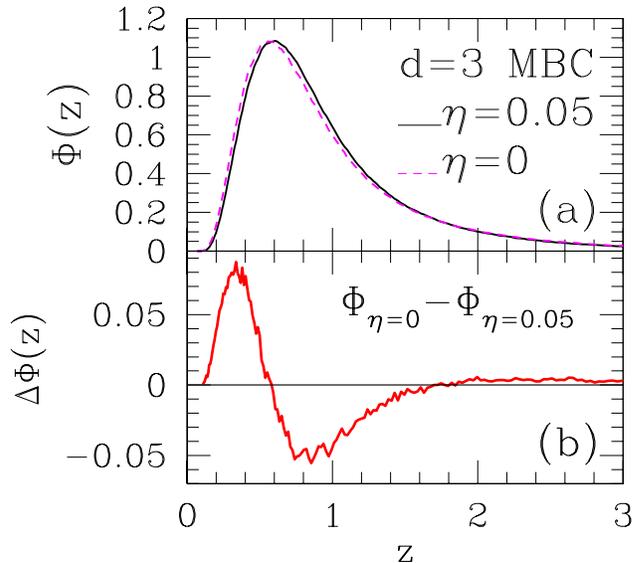}} \par}
\caption{(a) Scaled probability distributions $\Phi(z)$ in $d=3$ with MBC, 
for $z$ defined in Eq.~(\protect{\ref{eq:pvsphi}}).
Data for $L=40$, $5 \times 10^7$ configurations. Full line: 
demagnetization factor $\eta=0.05$; dashed line: $\eta=0$.
(b) Scaling function difference against $z$.
}
\label{fig:comp3df}
\end{figure}

We first investigated what could be learned about the
relevance of the self-regulating term, as regards roughness
distributions. In order to do so, we determined the approximate critical
value $H_e^c$ of the internal field $H_e$ of Eq.~(\ref{He}), by starting a
simulation with $\eta \neq 0$ and waiting for $H_e$ to stabilize. At that
point, we set $\eta=0$ and repeatedly varied $H$ in the interval
$(x\,H_e^c, H_e^c )$, $x \lesssim 1$, according to the procedure
delineated in Sec.~\ref{sec:2}. 
Though the interval of variation of $H$  did affect the size distribution 
of avalanches, as this is what characterizes the proximity of the 
depinning point~\cite{umm95,us}, no change was apparent in the  roughness 
data when comparing results, e.g., for $x=0.95$ and  $x=0.9$. 
For the simulations described in the remainder of this subsection, we used 
the latter value. In all cases studied, namely, $d=2$ PBC and $d=3$ with
both MBC and PBC, the influence of the demagnetization term on the roughness 
PDFs is rather small, but systematic.
This is illustrated in Fig.~\ref{fig:comp3df} for
$d=3$ with MBC, the case for which the deviations  between the $\eta \neq 
0$ and $\eta =0$ sets of data are the largest in magnitude. 
One sees that neglecting the demagnetizing term causes a small leftward
shift of the scaling curve. As we will see in Section~\ref{subsec:pbc},
the changes it causes to the fits of our distributions to the  analytical
$1/f^\alpha$ curves are of the order of systematic imprecisions
characteristic of the fitting procedure. Nevertheless, it is instructive
to seek the physical  origins of such effect. This is done by direct
inspection of the unscaled PDFs. 
\begin{figure}
{\centering \resizebox*{3.4in}{!}{\includegraphics*{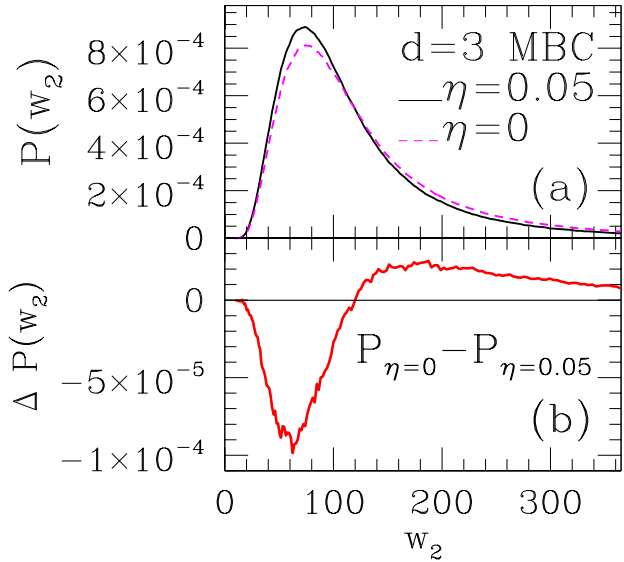}} \par}
\caption{(a) Probability distributions $P(w_2)$ in $d=3$ with MBC.
Data for $L=40$, $5 \times 10^7$ configurations. Full line: 
demagnetization factor $\eta=0.05$; dashed line: $\eta=0$.
(b) Probability distribution difference against $w_2$. Extent of 
horizontal axis corresponds to the same interval of $z$--variation
in Fig.~\protect{\ref{fig:comp3df}}.
}
\label{fig:pcomp3df}
\end{figure}
In Fig.~\ref{fig:pcomp3df} it is apparent that, for $\eta=0$ the 
high-end tail of $P(w_2)$ is slightly fatter than  for $\eta \neq 0$, at 
the  expense of a small amount of depletion around the most probable value 
of $w_2$. Accordingly, the average $\langle w_2 \rangle$ is higher by 
$\simeq 8\%$ in the former case than in the latter (the fractional difference 
between averages is the same also for $d=2$ and $d=3$ PBC). Such a trend 
can be understood by recalling that the $\eta=0$ data have been collected
just {\em below} the depinning transition, i.e., still within the
regime where pinning forces are dominant. Thus the interface mostly
meanders about, in order to comply with local energy minimization 
requests. The complement of this picture is that, for $H > H_c$
the interface moves with finite speed, more or less ignoring local 
randomness configurations, and becoming smoother the farther one is above
the critical point. In short, for a given lattice size the average 
interface roughness decreases monotonically as the external field
(driving force) is increased across its critical value. 

The interpretation of the small differences between $\eta=0$ and $\eta
\neq 0$ distributions is then as follows: (i) because of the way in which
data for the former were collected here, they represent a system just
below $H_c$, for which interface roughness is slightly larger than at the
critical point; and (ii) the closeness of $\eta=0$ data to those for $\eta
\neq 0$, and the way in which both sets of data differ, strongly suggest
that behavior {\it at} the critical point of the $\eta=0$ system is the
same as that of the $\eta \neq 0$ (self-regulated) case.  We conclude that
the self-regulating term is irrelevant, as far as critical roughness
distributions are concerned.
 
\subsection{PBC, $d=2$ and $3$}
\label{subsec:pbc}

Analytical expressions for the $1/f^\alpha$ distributions with PBC are 
either given in Ref.~\onlinecite{adgr02} ($d=2$), or can be derived
straightforwardly from Refs.~\onlinecite{adgr02,rkdvw03} ($d=3$).
In the latter case, the use of exact identities for two-dimensional 
lattice sums~\cite{zr75} speeds up calculations considerably.
Estimates of the exponent $\zeta$ of Eq.~(\ref{eq:zeta}), from 
power-law fits of simulational data with ${\cal O} (10^6)$ events, and 
$400 \leq L_x \leq 1200$  for $d=2$, $30 \leq L \leq 80$ for 
$d=3$, give $\zeta (d=2,{\rm PBC})= 1.24(1)$,
$\zeta (d=3,{\rm PBC}) = 0.71(1)$~\cite{bark3}.

Consideration of the scaling properties of height-height correlation 
functions and their Fourier transforms then suggests~\cite{rkdvw03},
for the generalized Gaussian case of independent Fourier modes, that
\begin{equation}
\alpha= d^\prime + 2\zeta\qquad (d^\prime=d-1)\ , 
\label{eq:alphazeta}
\end{equation}
which would imply $\alpha= 3.48(2)$ ($d=2),\ $3.42(2) ($d=3$). 

Such predictions can be
quantitatively checked by estimating the values of $\chi^2$ per degree of 
freedom ($\chi^2_{\rm \ d.o.f.}$) from fits of our simulation results to 
the analytical distributions. Since, even with $5 \times 10^7$ samples, 
the simulational data eventually get frayed at the top end, 
given the long forward tails characteristic of all systems studied here, 
our fits used only data for which $\Phi(z) \ge 10^{-3}$. This 
turned out not to be a drastic restriction, as we were left typically with
at least $100 - 200$ points to fit in each case.
\begin{figure}
{\centering \resizebox*{3.4in}{!}{\includegraphics*{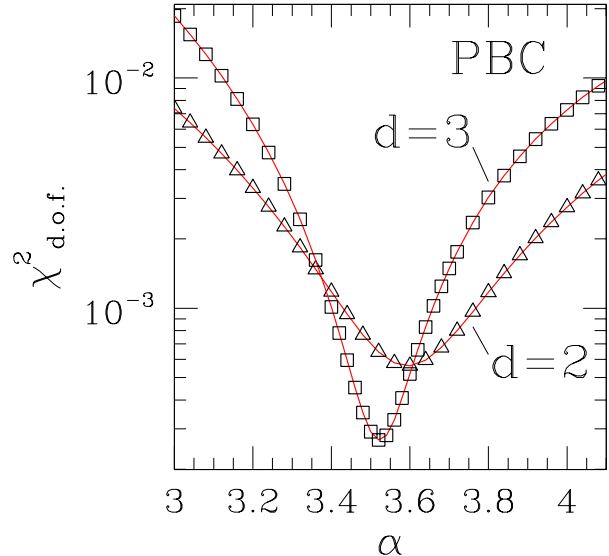}} \par}
\caption{ $\chi^2$ per degree of freedom ($\chi^2_{\rm \ d.o.f.}$)
for fits of simulation data with PBC to analytical forms of $1/f^\alpha$
distributions, against $\alpha$. Triangles: $d=2$, $L_x=400$;
squares, $d=3$, $L=40$.
}
\label{fig:chi2p}
\end{figure}
Assuming the uncertainty in the value of $\alpha$ that best fits our data 
to be given by requiring that $\chi^2_{\rm \ d.o.f.}$ stay within $150\%$ 
of its minimum, we quote from the data shown in  Fig.~\ref{fig:chi2p}: 
$\alpha =3.60(13)$ ($d=2$); $3.52(6)$ ($d=3$). The 
agreement with the above predictions is satisfactory, though slight 
discrepancies remain.
A visual check of the goodness-of-fit for each case is given in 
Figs.~\ref{fig:rdist2dp} and~\ref{fig:rdist3dp}. 

Fitting $\eta=0$ data to the closed-form distributions produces curves
whose minima of  $\chi^2_{\rm \ d.o.f.}$  are essentially the same as in
Fig.~\ref{fig:chi2p}, and slightly shifted rightwards. Using the same
criteria as above for the estimation of error bars, we have, for
$\eta=0$: $\alpha =3.64(16)$ ($d=2$); $3.59(5)$ ($d=3$). 

Detailed 
discussion, and pertinent comparisons with  data from 
Ref.~\onlinecite{rkdvw03}, will be deferred to Section~\ref{conc}.  
\begin{figure}
{\centering \resizebox*{3.4in}{!}{\includegraphics*{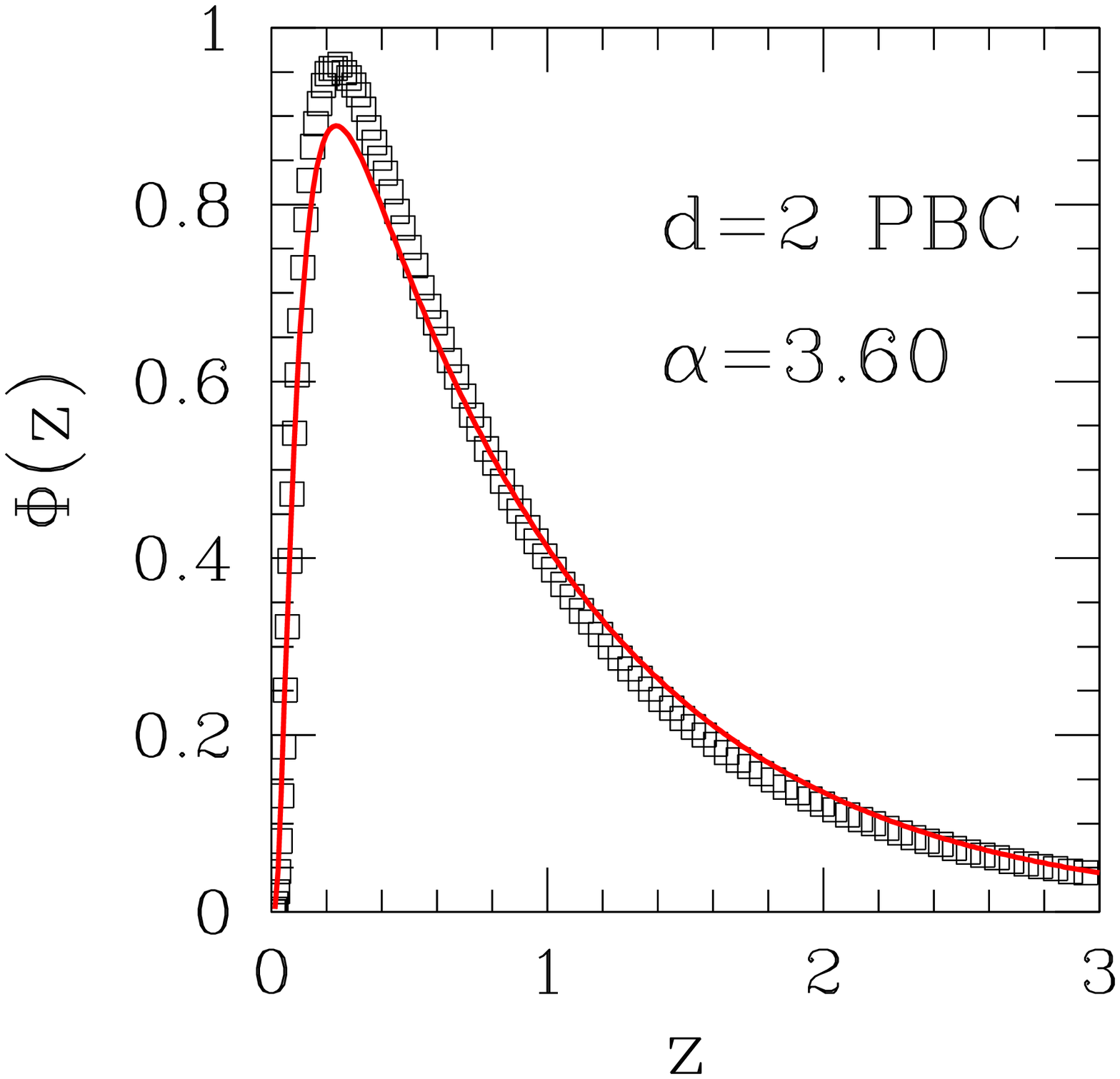}} \par}
\caption{Scaled probability distribution $\Phi(z)$ in $d=2$ (PBC), for 
$z$ defined in Eq.~(\protect{\ref{eq:pvsphi}}), from $5 \times 10^7$
configurations. Squares: simulation data ($L=400$). 
Full line is roughness distribution for $1/f^\alpha$ noise given in 
Ref.~\protect{\onlinecite{adgr02}}, with $\alpha=3.60$.
}
\label{fig:rdist2dp}
\end{figure}
\begin{figure}
{\centering \resizebox*{3.4in}{!}{\includegraphics*{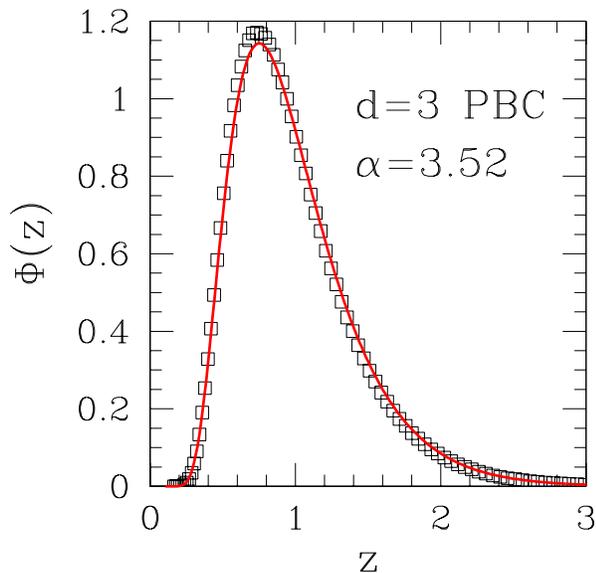}} \par}
\caption{Scaled probability distribution $\Phi(z)$ in $d=3$ with PBC, for 
$z$ defined in Eq.~(\protect{\ref{eq:pvsphi}}), from $5 \times 10^7$
configurations. Squares: simulation data ($L=40$). 
Full line is roughness distribution for $1/f^\alpha$ noise, with $\alpha=3.52$.
}
\label{fig:rdist3dp}
\end{figure}
 
\subsection{FBC and WBC, $d=2$ and $3$}
\label{subsec:fbc}

We have generated roughness data in both $d=2$ and $3$ with FBC.
Our initial implementation of FBC, used also in Ref.~\onlinecite{bark3}, 
aims at a literal reproduction of the constraint that the interface
must be horizontal at the edges. Thus, e.g. for $d=2$, ``ghost''
sites are added at $x=0$,~$x=L_x+1$, whose heights are always adjusted to 
be respectively $h(0,t)=h(1,t)$, $h(L_x+1,t)=h(L_x,t)$. This way, the edge 
sites at $x=1$ and $L_x$ experience no elastic pull (see the second 
term on the right-hand side of Eq.~(\ref{force})) from their ghost 
neighbors outside the sample.
 
Similarly to the PBC cases, estimates of the exponent $\zeta$ of 
Eq.~(\ref{eq:zeta}) were extracted from 
power-law fits of simulational data with ${\cal O} (10^6)$ events, and 
$400 \leq L_x \leq 1000$  for $d=2$, $30 \leq L \leq 80$ for 
$d=3$. The results are $\zeta (d=2,{\rm FBC})= 1.28(2)$,
$\zeta(d=3,{\rm FBC}) = 0.89(1)$. While the former value might be 
construed as not inconsistent with PBC and FBC giving the same 
universality class  for $d=2$, the same picture cannot hold for $d=3$. 
Though it is known~\cite{adgr01,adgr02,adr04,rkdvw03,bin81} 
that boundary conditions do have significant influence on
{\em scaling functions} of critical systems, they are not generally
expected to change the values of critical {\em exponents}.

In order to discuss the roughness PDFs, we first recall the effect of FBC 
on $1/f^\alpha$ distributions. The generating function
$G(s) =\int dw_2\,P(w_2)\,e^{-sw_2}$ has the general
form for PBC~\cite{adgr02,rkdvw03}
\begin{equation}
G_p(s) =\prod_{{\bf n}\neq 0}\left(1+\frac{s}{{\bf
n}^\alpha}\right)^{-1/2}\ ,
\label{eq:g(s)}
\end{equation}
where ${\bf n}$ is a lattice vector in $d-1$ dimensions with integer
coordinates. Because all ${\bf n}$ are counted, the square root disappears
due to the (at least) twofold degeneracy. Requiring that the interface be
horizontal at the edges implies that the Fourier representation of
$h(t)$ includes only cosines. The corresponding $G_f(s)$ has the
degeneracy of its singularities cut in half, compared to PBC. 

In $d=2$, this means that the single poles found for PBC turn into
square-root singularities.
Evaluation of $P(w_2)$, as the inverse Laplace transform of $G_f(s)$, 
thus necessitates a direct approach, since the residue theorem is 
inapplicable. This has been accomplished in Ref.~\onlinecite{mrkr04}, 
from which the relevant expressions were extracted in order to
attempt a minimization of $\chi^2_{\rm \ d.o.f}$ against 
$\alpha$, similar to that of Section~\ref{subsec:pbc}. 
With $\zeta (d=2,{\rm FBC})$ as above, one would expect a good fit
for $\alpha \simeq 3.5 - 3.6$. Instead, $\chi^2_{\rm \ d.o.f}$ has a 
minimum value $\simeq 4 \times 10^{-3}$  at $\alpha =2.96$,
and increases monotonically to reach $\simeq  4 \times 10^{-2}$ at $\alpha 
=3.5$. This is clearly at variance with correspondings results for
the PBC case. 

We then decided to generate data using window boundary conditions 
(WBC)~\cite{adgr02,mrkr04}, which are generally accepted as an alternative
way to simulate free edges. Accordingly, in $d=2$ 
we imposed global PBC on a system
of overall length $L_x$, and measured the local roughness within each 
of $n_w$ adjacent windows of length $L_x/n_w$. With $n_w \gg 1$,
it is plausible to assume that the resulting PDFs are independent of the 
boundary conditions established at $x=0$, $L_x$. In order 
to guarantee statistical independence, one should in principle use widely 
separated windows. However, the use of nonoverlapping, but 
neighboring, windows instead appears to introduce no measurable
errors on 
the resulting PDFs~\cite{adgr02}. We fixed $n_w=10$, and initially
measured $\zeta$ via Eq.~(\ref{eq:zeta}), from a sequence of
simulations with ${\cal O} (10^6)$ 
events (i.e. individual avalanches, thus the total number of roughness 
samples is larger by a factor of $n_w$), and $400 \leq L_x \leq 1200$, 
which gave $\zeta (d=2,{\rm WBC})= 1.21(2)$. Though this differs
by $3.5$ standard deviations from the value coming from FBC, it is
just consistent, at the margin, with $\zeta (d=2,{\rm PBC})= 1.24(1)$
found above.

\begin{figure}
{\centering \resizebox*{3.4in}{!}{\includegraphics*{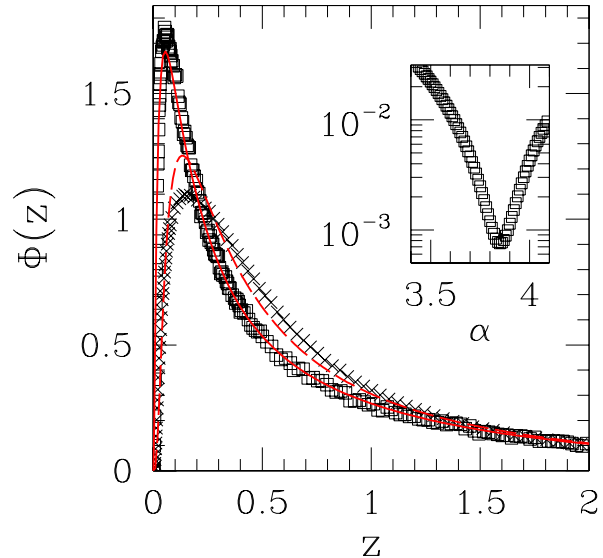}} \par}
\caption{Scaled probability distribution $\Phi(z)$ in $d=2$, for 
$z$ defined in Eq.~(\protect{\ref{eq:pvsphi}}).  Points are simulation
data. Crosses: $L=400$, FBC, $5 \times 10^7$ configurations.
Squares: $L=400$, WBC, $10^7$ avalanches, $n_w=10$  windows. Full line
is roughness distribution for $1/f^\alpha$ noise
(see Ref.~\protect{\onlinecite{mrkr04}}), with $\alpha=3.85$.
Dashed line : roughness distribution for $\alpha=2.96$ (see text). 
Inset: $\chi^2_{\rm \ d.o.f.}$ against $\alpha$, for fits of WBC 
simulation data against  $1/f^\alpha$ distributions, showing a minimum at
$\alpha=3.85$.
}
\label{fig:rdist2df}
\end{figure}

Direct examination of scaled PDFs results in the following observations.
First, in Figure~\ref{fig:rdist2df}  one can see that the PDFs in $d=2$ 
for FBC and WBC are unmistakably distinct. Furthermore, fits of FBC
data to the analytical expressions derived in Ref.~\onlinecite{mrkr04}
have been found to be generally of low quality. As mentioned above,
the best fit of FBC data is for the $\alpha = 2.96$ curve, shown in
the Figure as a dashed line, and corresponds to $\chi^2_{\rm \ d.o.f.}
\simeq 4 \times 10^{-3}$.
Though this average deviation is of the same order as that for the best 
case with PBC (recall Fig.~\ref{fig:chi2p}), 
comparison to Fig.~\ref{fig:rdist2dp} shows that, while for PBC
discrepancies are concentrated close to the narrow peak (thus they can be 
at least partially ascribed to binning effects), here one has a rather
widespread disagreement in shape.

On the other hand, WBC data can be much more closely fitted
by the analytical expressions, as shown both in the inset of 
Fig.~\ref{fig:rdist2df}, where $\chi^2_{\rm \ d.o.f.}$ 
exhibits a minimum  value $\simeq 7 \times 10^{-4}$ at $\alpha= 3.85$,
and directly in the main Figure, by the superposition of the $\alpha= 
3.85$ curve onto the corresponding numerical data.  

In summary, an analytical  form derived from assuming an interface
whose Fourier representation has only cosines (i.e. is horizontal at
the edges) has provided a very good fit to numerical data generated
by imposing WBC. Though this appears contradictory, the same
procedure  has been successfully accomplished in
Ref.~\onlinecite{mrkr04}, with regard to both experimental and
simulational data. 

Still, an important question remains, since the optimum $\alpha=3.85(5)$
(error bars estimated as in Section~\ref{subsec:pbc}) implies
$\zeta=1.43(3)$ via Eq.~(\ref{eq:alphazeta}). This is significantly
distinct from all three estimates thus far obtained for $d=2$, which
average to 1.25(5). We shall defer the discussion of this point to
Section~\ref{conc}.

Turning now to $d=3$, all poles of $G_p(s)$ have even degeneracy.
A straightforward adaptation for FBC is as follows. Recalling that
the lattice sums $\sum_{{\bf n}}|{\bf{n}}|^{-\alpha}$ which crop up
in the calculation of $\langle w_2 \rangle$~\cite{adgr01,adgr02,rp94}
must be halved, this implies a rescaling of the variable $s$, so
formally one can write~\cite{adgr02}:
\begin{equation}
G_f(s) =\sqrt{G_p(2s)}\ .
\label{eq:pvfs}
\end{equation}
Fitting our $d=3$ FBC data to analytical distribution
functions, obtained with help of Eq.~(\ref{eq:pvfs}), turns out to give
similar results to the $d=2$ case. 
The above-quoted value $\zeta=0.89(1)$, from the
finite-size scaling of $\langle w_2 \rangle$, together with
Eq.~(\ref{eq:alphazeta}), would suggest $\alpha= 3.78(2)$. However,
$\chi^2_{\rm \ d.o.f.}$ against $\alpha$ has a single
minimum ($\simeq 10^{-3}$) at $\alpha=3.18(8)$ (error bars estimated as in
Section~\ref{subsec:pbc}) and increases monotonicaly, reaching 
$\simeq 2 \times 10^{-2}$ at $\alpha= 3.78$. 

We again resorted to WBC. Imposing PBC at the edges of a system with $L 
\times L$ cross-section, we measured local roughness within each of 
$n_w$ non-overlapping, adjacent, square windows of linear dimension
$L/\sqrt{n_w}$ (or the largest integer contained in it). 
We took $n_w=16$, and initially  measured $\zeta$ from a sequence of 
simulations with ${\cal O} (10^6)$ events, and $30 \leq L \leq 80$, 
which gave $\zeta (d=3,{\rm WBC})= 0.75(2)$. The discrepancy between this
and the value $0.89(1)$ coming from FBC is rather more severe than the 
corresponding case for $d=2$. On the other hand, the present estimate
is close to  the values of $\zeta (d=3,{\rm PBC})$ found above, namely
$0.71(1)$ from Eq.~\ref{eq:zeta}, and $0.76(3)$ from optimization
of fits against $1/f^\alpha$ distributions plus
Eq.~\ref{eq:alphazeta}. 

\begin{figure}
{\centering \resizebox*{3.4in}{!}{\includegraphics*{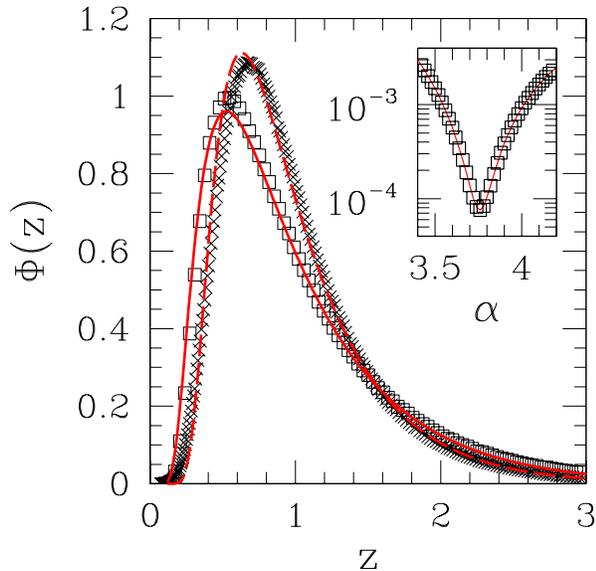}} \par}
\caption{Scaled probability distribution $\Phi(z)$ in $d=3$, for 
$z$ defined in Eq.~(\protect{\ref{eq:pvsphi}}). Points are
simulation data. Crosses: $L=40$, FBC, $5 \times 10^7$ 
configurations. Squares: $L=40$, WBC, $3 \times 10^7$ avalanches,
$n_w=16$ windows. Lines are
roughness distributions for $1/f^\alpha$ noise (see 
Eq.~(\protect{\ref{eq:pvfs}})), with
$\alpha=3.76$ (full) and $3.18$ (dashed). 
Inset: $\chi^2_{\rm \ d.o.f.}$ against $\alpha$, for fits of
WBC simulation data against  $1/f^\alpha$ distributions, showing a
minimum at $\alpha=3.76$.
}
\label{fig:rdist3df}
\end{figure}

Again, we investigated the roughness PDFs generated with WBC. 
Similarly to the  $d=2$  case, they differ markedly from the ones
obtained with FBC, as 
shown in Fig.~\ref{fig:rdist3df} . This time, fits against the analytical 
expressions given through Eq.~(\ref{eq:pvfs}) exhibit a deep, well-defined 
minimum of $\chi^2_{\rm \ d.o.f.}$ at $\alpha=3.76(5)$ (see inset in the
Figure), in very good agreement with $\alpha= 3.78(2)$ predicted from  
finite-size scaling of $\langle w_2 \rangle$ data for FBC, together with
Eq.~(\ref{eq:alphazeta}). However, for reasons to be explained at length
in Section~\ref{conc}, we believe this coincidence to be accidental.
 
\subsection{MBC, $d=3$}
\label{subsec:mbc}

We started by studying systems with a square cross-section, imposing 
PBC along $x$ and FBC, as defined at the beginning of
Section~\ref{subsec:fbc}, along $y$.
 
Estimates of the exponent $\zeta$ of Eq.~(\ref{eq:zeta}) were 
again extracted
from power-law fits of simulational data with ${\cal O} (10^6)$ events, and 
$30 \leq L \leq 80$ for $d=3$ MBC, with the result
$\zeta(d=3,{\rm MBC}) = 0.87(1)$~\cite{bark3}. 

The Fourier representation of $h(t)$ with MBC can be put in the form:
\begin{equation}
h(x,y) = \sum_{m,n} c_{mn}\,e^{2\pi i\left(mx
+\frac{n}{2}y\right)/L}\ ,
\label{eq:mbc}
\end{equation}
where $m,n= 0,\pm 1, \pm 2, \dots$, $(m,n) \neq (0,0)$, and $c_{-m,n} =
c^\ast_{m,n}$; $c_{m,-n}=c_{m,n}$. Thus a global rescaling such as that of
Eq.~(\ref{eq:pvfs}) is not possible. On the other hand, starting from 
Eq.~(\ref{eq:mbc}), an analysis
similar to that of Refs.~\onlinecite{adgr02,rp94} suggests a
generating function:
\begin{equation}
G_m(s) =\prod_{m,n}\left(1+\frac{s}{(
4m^2+n^2)^{\alpha/2}}\right)^{-1/2}\ ,
\label{eq:gm(s)}
\end{equation}
again with $(m,n)\neq (0,0)$. The double sum
$\sum_{m,n} (4m^2+n^2)^{-\alpha/2}$, which appears in the subsequent
expression for $\langle w_2 \rangle$, corresponds to ${\cal Q}(1,0,4)$
of Ref.~\onlinecite{zr75} and can be easily evaluated.

We performed fits of simulational data to the closed-form PDFs
calculated as above.
While  Eq.~(\ref{eq:alphazeta}), with $\zeta=0.87(1)$, gives 
$\alpha=3.74(2)$, $\chi^2_{\rm \ d.o.f}$  has a minimum $\simeq 2
\times 10^{-3}$ at $\alpha=3.36(10)$. The overall quality of fits is
slightly worse than for $d=3$ FBC (refer to Fig.~\ref{fig:rdist3df}).

In order to investigate WBC, we took rectangular systems with 
dimensions $L_x$ and  $L_y=4\,L_x$
with full PBC  (we denote this setup as {\em mixed window} boundary 
conditions (MWBC)) and calculated local roughness distributions within
$n_w=4$ square windows of $L_x \times L_x$ sites each, side by side along 
the $y$ axis. Scaling of the first moment of the distribution, 
Eq.~(\ref{eq:zeta}), with $30 \leq L_x \leq 80$, gave $\zeta = 0.74(1)$.

Again, the roughness PDF thus obtained was markedly distinct
from that with MBC. In addition, fits to the analytical
expressions derived from Eq.~(\ref{eq:gm(s)}) were considerably worse
than those of MBC data, with a minimum $\chi^2_{\rm \ d.o.f} \simeq 1
\times 10^{-2}$ at $\alpha=4.1$. 

\begin{figure}
{\centering \resizebox*{3.4in}{!}{\includegraphics*{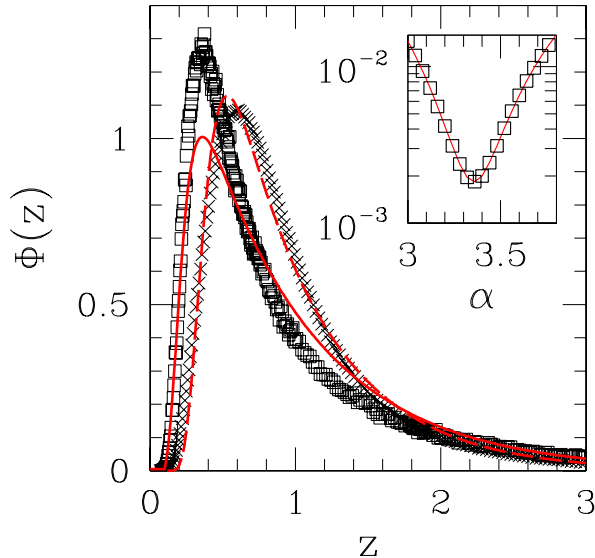}} \par}
\caption{Scaled probability distribution $\Phi(z)$ in $d=3$, for 
$z$ defined in Eq.~(\protect{\ref{eq:pvsphi}}).
Crosses: simulation data ($L=40$, MBC, $5 \times 10^7$
configurations).
Squares: simulation data ($L_x=40$, $L_x=160$, MWBC (see text), $3
\times 10^6$ avalanches, $n_w=4$ windows). Lines are
roughness distributions for $1/f^\alpha$ noise (see 
Eq.~(\protect{\ref{eq:gm(s)}})), with
$\alpha=4.1$ (full) and $3.36$ (dashed). 
Inset: $\chi^2_{\rm \ d.o.f.}$ against $\alpha$, for fits of
MBC simulation data against  $1/f^\alpha$ distributions, showing a
minimum at $\alpha=3.36$.
}
\label{fig:rdist3dm}
\end{figure}

The results are displayed in Fig.~\ref{fig:rdist3dm}, where it can be 
seen that even the best-fitting analytical PDF fails to provide a
good match to the MWBC data (except for the initial, rather steep,
ascent close to $z=0$).
    
\section{Discussion and Conclusions}
\label{conc}
We begin our discussion by recalling from
Ref.~\onlinecite{bark3} and Sec.~\ref{subsec:pbc} that, for the model
considered here with PBC, the finite-size scaling of the first moment of
the distribution gives $\zeta (d=2,{\rm PBC})= 1.24(1)$,
$\zeta(d=3,{\rm PBC}) = 0.71(1)$. Both compare  well with the usually
accepted values for the quenched Edwards-Wilkinson (EW) universality
class~\cite{les93,ma95,mbls98,rhk03},
respectively $\zeta \simeq 1.25$ ($d=2$) and $\zeta \simeq 0.75$ ($d=3$).
Furthermore, consideration of the full distributions points the same
way: our simulational data displayed in Figs.~\ref{fig:rdist2dp} 
and~\ref{fig:rdist3dp} match very well those in Figure~2
of Ref.~\onlinecite{rkdvw03} which concern the EW model.
The agreement  with EW behavior is consistent with our results of
Sec.~\ref{subsec:eta0} regarding
the independence of scaled roughness distributions on the demagnetizing
term. Indeed, the quenched  EW equation can be written as~\cite{rhk03}
\begin{equation}
\frac{\partial h({\bf x},t)}{\partial t} =u({\bf x},h)+ a\,\nabla^2 
h({\bf x}) + f\ ,
\label{eq:EW}
\end{equation}
where $u$ represents quenched disorder and $f$ is the external driving
force. This has a one-to-one correspondence with Eq.~(\ref{force}),
except that in that Equation we allowed for the self-regulating,
demagnetizing, term. Having shown that
such  mechanism is irrelevant as far as scaled roughness distributions
are concerned, it becomes tenable to assume that, overall,
our model belongs to the EW universality class. 

Still for PBC, the connection between the exponents $\alpha$ and $\zeta$,
predicted~\cite{rkdvw03} in Eq.~(\ref{eq:alphazeta}), is verified  within
reasonable error bars. 


Turning to different sets of boundary conditions, we first point
out that small differences in implementation of FBC (namely,
``literal'' FBC, i.e. horizontal interface at the edges, versus WBC)  
significantly alter the roughness PDFs.  The question then arises
of which, if any, of these implementations is the ``right'' one.

We investigate this by referring to results derived through a ``proven'' 
method, i.e.
finite-size scaling of the first moment of the distribution.
Examination of the corresponding column of Table~\ref{table1} strongly 
suggests that, in both $d=2$ and $3$, WBC (including WMBC) preserves 
universality  with PBC, while FBC does not (though in $d=2$ FBC does not 
perform very  badly). Accepting such preservation as a basic tenet, we 
conclude  that FBC as implemented induces strong distortions in the 
scaling behavior  of interface roughness. In this context, the
good agreement in $d=3$ between the optimum $\alpha$ for fits of WBC data 
to the analytical forms, and that coming from finite-size scaling of FBC 
data via Eqs.~(\ref{eq:zeta})~and~(\ref{eq:alphazeta}), must be regarded 
as fortuitous.

Thus, we discard FBC, as well as MBC, for the remaining of the present
discussion. One must note, however, that use of MBC (i.e. partial FBC)  
provides a sensible representation of the physical setup found in thin
films, as well as reproducing well-known results (concerning scaling
behavior of avalanche sizes) at both ends of the crossover between $d=2$
and $3$~\cite{bark3}.

\begin{table}
\caption{ Estimates of roughness exponent $\zeta$ for different 
dimensionalities and boundary conditions (BC). $\zeta^{\rm FSS}$:
finite-size scaling of first moment of distribution, 
Eq.~(\protect{\ref{eq:zeta}}). $\zeta_{\rm fit}$: from best-fitting
$1/f^\alpha$ distribution and Eq.~(\protect{\ref{eq:alphazeta}}).
$\chi^2_{\rm \ d.o.f.}$ (min): value of $\chi^2_{\rm \ d.o.f.}$
for $\zeta=\zeta_{\rm fit}$. 
}
\vskip 0.1cm
 \halign to \hsize{\hskip1.5truecm\hfil#\hfil&\hfil#\hfil&\hfil#\hfil&
\hfil#\hfil\cr
    \  &\ $\zeta^{\rm FSS}$ &\ $\zeta_{\rm fit}$&\ $\chi^2_{\rm \ 
d.o.f.}$ (min) \cr
  $d=2$ PBC &\ $1.24(1)$  &\ $1.30(8)$ &\ $6 \times 10^{-4}$\cr
  $d=2$ FBC &\ $1.28(2)$  &\ $0.98(7)$ &\ $4 \times 10^{-3}$\cr
  $d=2$ WBC &\ $1.21(2)$  &\ $1.42(3)$ &\ $7 \times 10^{-4}$\cr
  $d=3$ PBC &\ $0.71(1)$  &\ $0.76(3)$ &\ $3 \times 10^{-4}$\cr
  $d=3$ FBC &\ $0.89(1)$  &\ $0.59(4)$ &\ $1 \times 10^{-3}$\cr
  $d=3$ WBC &\ $0.75(2)$  &\ $0.88(1)$ &\ $8 \times 10^{-5}$\cr
  $d=3$ MBC &\ $0.87(1)$  &\ $0.68(5)$ &\ $2 \times 10^{-3}$\cr
  $d=3$ MWBC &\ $0.74(1)$  &\ $1.05(10)$ &\ $1 \times 10^{-2}$\cr}
\label{table1}
\end{table}
Returning to roughness scaling, we see in Table~\ref{table1} that the fair 
agreement between
$\zeta^{\rm FSS}$ and $\zeta_{\rm fit}$, found for PBC in $d=2$ and $3$,
is absent in the remaining cases under consideration, i.e. $d=2$ WBC, 
$d=3$ WBC, $d=3$ MWBC.
One might ask whether finite-size effects (though widely believed to 
vanish already for small lattices~\cite{adgr01,adgr02,rkdvw03,forwz94}) 
still have a nonnegligible quantitative effect  on the scaled roughness 
PDFs found here, so as to distort our fits to the analytical 
distributions. We present data to show that this is not the case.

\begin{figure}
{\centering \resizebox*{3.4in}{!}{\includegraphics*{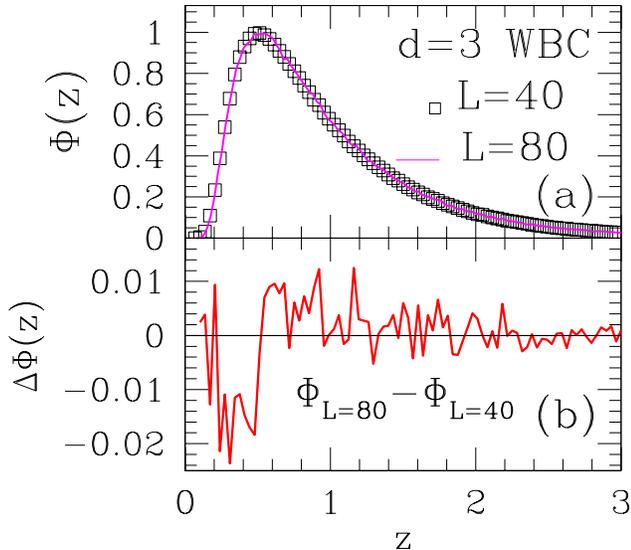}} \par}
\caption{(a) Scaled probability distributions $\Phi(z)$ in $d=3$ with WBC, 
for $z$ defined in Eq.~(\protect{\ref{eq:pvsphi}}).
Squares: $L=40$. Full line: $L=80$. In both cases, $10^6$ avalanches, 
$n_w=16$ windows. (b) Scaling function difference against $z$.
}
\label{fig:phidiff}
\end{figure}

In Figure~\ref{fig:phidiff} we compare $L=40$ and $L=80$ PDFs, for $d=3$ 
WBC. Contrary to the systematic trend exhibited in Fig.~\ref{fig:comp3df} 
(for comparison between $\eta=0$ and $\neq 0$ distributions), here the 
difference $\Delta\Phi(z)$ is rather small and essentially random, arising 
because of fluctuations in statistics, coupled with binning effects. 
An apparently systematic effect shows up only for the narrow range close 
to $z=0$ where both PDFs have a steep slope. That, however, involves
only of order $5-10$ points, with a consequently reduced effect on the
overall statistics. The corresponding curves $\chi^2_{\rm \ d.o.f.}$
against $\alpha$ are nearly indistinguishable; with $L=80$ data, the 
minimum of $\chi^2_{\rm \ d.o.f.}$ is $9 \times 10^{-5}$ at $\alpha= 
3.76(4)$, virtually identical to the $L=40$ result shown in 
Fig~\ref{fig:rdist3df} (see also Table~\ref{table1}). For $d=2$ WBC
and $d=3$ MWBC, the overall picture is the same.
Therefore, finite-size effects on the numerically-obtained PDFs are not a 
likely source for the disagreements found.    

We note also that, when considering $1/f^\alpha$ distributions, there is no
apparent reason why Eq.~(\ref{eq:alphazeta}) should not hold for
boundary conditions other than PBC, as that Equation was derived for
generalized Gaussian distributions~\cite{rkdvw03} with the only
assumption that the large-scale behavior is determined by a single
observable.

We are thus left with a single point to analyze, namely the overall
adequacy of $1/f^\alpha$ distributions to describe the problem at hand. 
The following comments are in order:

(1) already for PBC, the study of generalized depinning problems shows
that small but systematic discrepancies remain between numerical data and
$1/f^\alpha$ PDFs, whose origins can be traced to higher cumulants of the
correlation functions~\cite{rkdvw03}. Thus, in this sense the $1/f^\alpha$
distributions are not expected to be a perfect fit, even for PBC.

(2) In Ref.~\onlinecite{mrkr04} the equation of motion for $h(x)$ contains
a long-range elastic term, $\int
dx_1\,\left(h(x)-h(x_1)\right)/(x-x_1)^2$, instead of the local term,
$\nabla^2 h({\bf x})$, present here. While in that case an $1/f^\alpha$
distribution gives good fits to the numerically-generated roughness PDF
with WBC, this does not necessarily imply that a similar quality of fit
can be found for the present EW problem with WBC. In this connection, one
might ask how far the independent Fourier mode assumption, basic in the
derivation of $1/f^\alpha$ PDFs, is affected by such details. One sees
that the long-range term contributes qualitatively in the same direction
as PBC, i.e. by imposing additional constraints on interface roughness
(when compared, respectively, to short-range interactions and WBC).

A plausible scenario then emerges, in which the amplitude of corrections
to the representation of an interface roughness PDF by an $1/f^\alpha$ 
distribution would depend on how much that interface is constrained,
either by boundary conditions or by elastic terms in the equation of 
motion. Lessening of such constraints would imply an increase in
the correction amplitudes. However, at present we do not see a way to
quantify and test these remarks.

Clearly, more work is needed in order to clarify
the connection between $1/f^\alpha$ distributions and generalized
depinning transitions.

\begin{acknowledgments}

The author thanks Tibor Antal and Zolt\'an R\'acz for their
advice on numerical evaluation of the closed-form PDFs, 
as well as Robin Stinchcombe and J. A. Castro for interesting discussions 
and suggestions. Thanks are also due to a referee for pointing out 
Ref.~\onlinecite{mrkr04}. This research  was partially supported by 
the Brazilian agencies CNPq (Grant No. 30.0003/2003-0), FAPERJ (Grant
No. E26--152.195/2002), FUJB-UFRJ and Instituto do Mil\^enio de
Nanoci\^encias--CNPq.
\end{acknowledgments}

\bibliography{biblio}  
\end{document}